\documentclass[journal=jacsat,manuscript=article]{achemso}

\usepackage[version=3]{mhchem} 

\author{M.~Saad Bin-Alam}
\affiliation{School of Electrical Engineering and Computer Science, University of Ottawa, Ottawa, ON \ K1N 6N5, Canada \\}
\author{Joshua~Baxter}
\affiliation{Department of Physics, University of Ottawa, Ottawa, ON \ K1N 6N5, Canada \\}
\author{Kashif M.~Awan}
\alsoaffiliation{School of Electrical Engineering and Computer Science, University of Ottawa, Ottawa, ON \ K1N 6N5, Canada \\}
\affiliation{Stewart Blusson Quantum Matter Institute, University of British Columbia, Vancouver, BC, V6T 1Z4, Canada \\}
\author{Antti~Kiviniemi}
\affiliation{Laboratory of Photonics, Tampere University, FI-33100 Tampere, Finland \\}
\author{Yaryna~Mamchur}
\alsoaffiliation{School of Electrical Engineering and Computer Science, University of Ottawa, Ottawa, ON \ K1N 6N5, Canada \\}
\affiliation{National Technical University of Ukraine, “Igor Sikorsky Kyiv Polytechnic Institute,” Kyiv, Ukraine \\}
\author{Antonio Cal\`a Lesina}
\affiliation{Department of Physics, University of Ottawa, Ottawa, ON \ K1N 6N5, Canada \\}
\alsoaffiliation{School of Electrical Engineering and Computer Science, University of Ottawa, Ottawa, ON \ K1N 6N5, Canada \\}
\alsoaffiliation{Hannover Centre for Optical Technologies and Cluster of Excellence PhoenixD, Leibniz University Hannover, 30167, Hannover, Germany \\}
\author{Kosmas L.~Tsakmakidis}
\affiliation{Section of Condensed Matter Physics, Department of Physics, National and Kapodistrian University of Athens, Panepistimioupolis, GR-157 84 Athens, Greece}
\author{Mikko~J.~Huttunen}
\affiliation{Laboratory of Photonics, Tampere University, FI-33014 Tampere, Finland \\}
\author{Lora Ramunno}
\affiliation{Department of Physics, University of Ottawa, Ottawa, ON \ K1N 6N5, Canada \\}
\author{Ksenia~Dolgaleva}
\affiliation{School of Electrical Engineering and Computer Science, University of Ottawa, Ottawa, ON \ K1N 6N5, Canada \\}
\alsoaffiliation{Department of Physics, University of Ottawa, Ottawa, ON \ K1N 6N5, Canada \\}
\email{ksenia.dolgaleva@uottawa.ca}

\title[An \textsf{achemso} demo]
  {Hyperpolarizability of plasmonic meta-atoms in metasurfaces}

\abbreviations{IR,NMR,UV}

\begin{document}







\begin{abstract}
Plasmonic metasurfaces are promising as enablers of nanoscale nonlinear optics and flat nonlinear optical components. Nonlinear optical responses of such metasurfaces are determined by the nonlinear optical properties of individual nanostructured plasmonic meta-atoms, which are the building blocks of the metasurfaces. Unfortunately, no simple methods exist to determine the nonlinear coefficients (hyperpolarizabilities) of the meta-atoms hindering designing of nonlinear metasurfaces. Here, we develop the equivalent $RLC$ circuit model of such meta-atoms to estimate their second-order nonlinear optical parameter $i.e.$ the first-order hyperpolarizability in the optical spectral range. In parallel, we extract from second-harmonic generation experiments the spectrum of the 1$^\text{st}$-order hyperpolarizabilities of individual meta-atoms consisting of asymmetrically shaped (elongated) plasmonic nanoprisms. Moreover, we verify our results using nonlinear hydrodynamic-FDTD and with calculations based on nonlinear scattering theory. All three approaches: analytical, experimental, and computational, yield results that agree very well. Our empirical $RLC$ model can thus be used as a simple tool to enable efficient design of nonlinear plasmonic metasurfaces.
\end{abstract}


\noindent Photonic metamaterials are artificial 2D structures exhibiting optical properties that in natural materials are either very weak or entirely lacking. Among these properties are optical magnetism, strong chirality and epsilon-near-zero behavior~\cite{Alu2007, Zhang2009, Soukoulis2011}. In addition, there is a growing interest in understanding and harnessing the nonlinear optical responses of metasurfaces~\cite{Kauranen2012review, Lapine2014, ButetReview2015, Lesina2017, Li2017b, Rahimi2018}. This is due to the fact that many photonic applications including frequency conversion, ultrashort-pulse generation, photon-pair generation, all-optical switching and frequency-comb generation~\cite{Kwiat1995, Brabec2000, Kippenberg2011comb, Shcherbakov2015} rely on nonlinear optics occurring in large bulky devices where one must contend with phase mismatching. In contrast, the small footprint of metasurfaces virtually guarantees phase matching, and moreover, the nonlinear emission can be precisely controlled~\cite{Alu2007}.  

Nonlinear plasmonic metasurfaces have recently emerged as a promising candidate for enabling nanoscale nonlinear optics~\cite{Kauranen2012review}. The optical responses of plasmonic meta-atoms serving as unit cells of metasurfaces are dictated by the collective movement of the conduction electrons giving rise to localized surface plasmons (LSPs). Therefore, it is imperative to investigate the conduction electron dynamics and the nonlinear response of the constituting meta-atoms. Such investigations can be performed, for example, by using the hydrodynamic plasma model~\cite{Sipe1980,scalora_second-_2010,Ciraci2012,Ginzburg2015}, and the nonlinear scattering theory~\cite{Obrien2015}. However, relying on computational tools is not always convenient due to their complexity and the large amount of computational resources they often require. 

Linear and nonlinear optical responses of plasmonic meta-atoms can also be predicted by using a simpler model based on an equivalent $RLC$ circuit theory~\cite{Engheta2005, Huang2009, Zhou2005, Tretyakov2007, Staffaroni2012}. This approach has been demonstrated to correctly describe the nonlinear and magnetic responses of split-ring resonators operating at microwave frequencies~\cite{Poutrina2010}. Nevertheless, it remained unclear whether the $RLC$ approach could describe conduction electron dynamics occurring in plasmonic meta-atoms adequately enough to allow for accurate predictions of their nonlinear responses at optical wavelengths as well. 
In this letter, we empirically derive the first-order hyperpolarizability of individual plasmonic meta-atoms by adapting the equivalent $RLC$ model that we find can, indeed, be used to easily and quickly predict the nature of collective second-order nonlinear responses of large metasurface arrays. First, we derive an expression for the first-order hyperpolarizability for an unknown nonlinear coefficient $a$ that represents the strength of the nonlinear charge oscillation in an individual meta-atom. The goal of this work is to determine the value of this unknown coefficient through physical arguments, and to validate this approach through rigorous experiments and numerical calculations. Next, we describe our second-harmonic generation (SHG) experiments, where we measured SHG emission from metasurfaces consisting of randomly arranged gold elongated nanoprism meta-atoms. From these measurements, we extract the spectrum of their first-order hyperpolarizability. This experimental result is then validated by two sets of finite-difference time domain (FDTD) calculations. In the first set, the material nonlinearities are directly implemented within the code via a nonlinear hydrodynamic plasma model. The second set is based on nonlinear scattering theory where the experimentally-determined second-order nonlinear permittivity of gold is used.

Finally, we present a simple empirical derivation for the unknown $RLC$-model nonlinear coefficient $a$, which we then use to perform an order-of-magnitude estimate of the first-order hyperpolarizability spectrum. Our simple intuitive physical interpretation of the nonlinear parameter $a$ is such that the $RLC$ circuit model we present can be extended to describe metasurfaces with different shapes of meta-atoms without requiring a-priori experimental validation. This simple analytical tool, capable of predicting the values of hyperpolarizabilities of metamolecules based on the information about their shapes, dimensions and material compositions, will prove indispensable in the realization of metasurfaces with tailored nonlinear optical responses.

\begin{figure}[tbp]
\centering
{\includegraphics[width=0.80\linewidth]{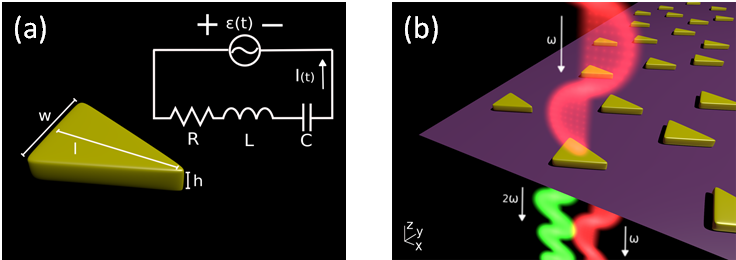}}
\caption{(a) An artist's depiction of an equivalent $RLC$ circuit diagram for the investigated nanoprisms. The time-varying incident field $\tilde{E}_{\rm{inc}}$ gives rise to the current of conduction electrons $\tilde{I}$, that is damped by electron collisions. (b) A metasurface consisting of a random array of triangular gold nanostructures.} 
\label{fig:RLCschematic}
\end{figure}

We begin by introducing the nonlinear $RLC$ model, which is schematically illustrated in Fig.~\ref{fig:RLCschematic}(a). We consider an incident field $\tilde{E}_{\rm{inc}}=E_0 \exp (-i \omega t)$, linearly polarized parallel to the length $l$ of the meta-atom, and assume that the field drives the $RLC$ circuit by creating an electromotive force $\tilde{\varepsilon}$ and, subsequently, a current $\tilde{I}= \dot {\tilde{q}}$ of conduction electrons with charge $q$. Throughout the text, the tilde and the over-dot notations denote time-varying quantities and time derivatives, respectively. The dynamics of the current, damped by electron collisions, can be described by the equation~\cite{Poutrina2010}   
\begin{equation}  \label{Eq:DE_current}
	L\dot {\tilde{I}} + R\tilde{I} + \tilde{V}_C(\tilde{q}) 
    		= \tilde{\varepsilon} (t) \, ,   
\end{equation}
where $L$ is the distributed inductance, $R$ is the distributed resistance, and $\tilde{V}_C$ is the induced voltage due to the effective capacitance $C$ of the circuit. $\tilde{V}_C$ is, in general, a nonlinear function of the charge $\tilde{q}$. Here, we assume that the nonlinearity of the system is sufficiently weak to allow us to write the nonlinear voltage function as a second-order polynomial~\cite{Poutrina2010} 
\begin{equation}
\label{Eq:V_C}
V_C(\tilde{q})=(\tilde{q}+a\tilde{q}^2)/C.
\end{equation}
This assumption is valid for at least up to ${\sim}10$~MW/cm$^2$ which is consistent with the experiments we have performed.

The specific form of the electromotive force $\tilde{\varepsilon} (t)$ depends on the oscillation direction of the incident field with respect to the meta-atom. In case of a nanorod-like meta-atom, if we assume that the incident field is polarized along its length $l$, the electromotive force $\tilde{\varepsilon} (t)$ takes the form $\tilde{\varepsilon}=\tilde{E}_{\rm{inc}} l$, assuming that the interaction of the meta-atom as a whole with the incident field can be treated as an electric dipole interaction. After substituting this expression into Eq.~(\ref{Eq:DE_current}), we obtain

\begin{equation}  \label{Eq:DE_charge_nonlinear}
	\ddot {\tilde{q}} + 2 \gamma \dot {\tilde{q}} + \omega_0^2 \tilde{q} +a\omega_0^2 \tilde{q}^{2} 
    		= C \omega_0^2 l \tilde{E}_{\rm{inc}} \, , 
\end{equation}
where $\omega_0 = 1/\sqrt{LC}$ is the resonance frequency of the circuit and $\gamma= R/2L$ 
is the free-electron damping constant of gold~\cite{Poutrina2010}. These relations take into account the dispersive nature of gold by implicitly considering a Drude model with $\gamma$ in Eq.~(\ref{Eq:DE_charge_nonlinear}) and $\omega_\mathrm{p}$ in Eq.~(\ref{Eq:Inductance_rect}) given in the Methods section~\cite{Poutrina2010}. 

When the excitation wavelengths are close to the LSP resonance of the
meta-atom, the conduction electron dynamics of the system is well described using the $RLC$ approach~\cite{Huang2009}. Eq.~(\ref{Eq:DE_charge_nonlinear}) is the master equation describing the dynamics of the conduction electrons, and therefore, the linear and nonlinear optical response of the meta-atom. A  steady-state solution to Eq.~(\ref{Eq:DE_charge_nonlinear}) can be found by implementing perturbation theory \cite{Boyd2020}, and is given by
\begin{equation}
\label{Eq:NL_perturbation}
   \tilde{q} = \frac{C \omega_0^2 l}{D(\omega)}E_{\rm{inc}} e^{-i\omega t} 
   - \frac{a C^2 \omega_0^6 l^2}{{D}^2 (\omega){D} (2\omega)}  E_{\rm{inc}}^2 e^{-i2\omega t} \, , 
\end{equation} 
where $D(\omega')=(\omega_{0}^2-\omega'^2 -2 i \gamma \omega')$. Next, we write the total induced dipole moment as $\tilde{p}=\tilde{q} l$, and 
recall that the dipole moment can be expressed in the frequency domain as~\cite{Boyd2020}
\begin{equation}  \label{Eq:polarizability} 
  	p = \epsilon_0 \alpha E_{\rm{inc}} 
        + \epsilon_0 \beta E^2_{\rm{inc}}  \, ,
\end{equation} 
where $\alpha$ is the linear polarizability and $\beta$ is the first-order hyperpolarizability of the meta-atom. We can now combine Eqs.~(\ref{Eq:NL_perturbation}) and~(\ref{Eq:polarizability}) to obtain simple equations for $\alpha$ and $\beta$:    
\begin{subequations}
\label{Eq:polarizabilities}   
  \begin{eqnarray}
    \label{Eq:alpha}   
    \alpha(\omega) &=& -\frac{C \omega_0^2 l^2}{\epsilon_0 D (\omega)}   \, , \\
    \label{Eq:beta}
     \beta (2\omega;\omega) &=& -a\frac{C^2 \omega_0^6 l^3}{\epsilon_0 D (2\omega){D^2 (\omega)}}   \, . 
  \end{eqnarray} 
\end{subequations}

As the lengths of the meta-atoms can be a significant fraction of the exciting wavelength, higher-order multipoles can be excited, and one must consider this possibility in general, especially for nonlinear emission at shorter wavelengths~\cite{Butet2010}. However, in this study, our goal is to create a model that gives an order of magnitude estimate of SHG generation, through the determination of the nonlinear coefficient $a$. Thus, to simplify our approach here, we consider only the electric dipole term, and therefore the hyperpolarizability in Eq.~(\ref{Eq:beta}) can be considered an ``effective'' hyperpolarizability that helps us to achieve this goal. 

In the following, we seek to determine the nonlinear coefficient $a$. In order to find its value empirically, we first extract its value from performed experiments, and then validate the result by comparing it against predictions based on two distinct implemented  computational methods.

SHG is a coherent second-order nonlinear, and therefore, it is very sensitive to the symmetry properties of the material under investigation~\cite{Boyd2020}. In fact, centrosymmetric-shaped structures $e.g.$ rectangular nanobars exhibit very weak SHG responses that are allowed under the electric dipole approximation of the light-matter interaction~\cite{Czaplicki2018} (see also Fig.~\ref{FIG:S1} in the supplementary information S1). In such centrosymmetric structures, the weak SHG response originates due to symmetry breaking at side surfaces of the nanobars. As a result, nanostructures with low symmetry, such as split-ring resonators, L-shapes or nanoprisms, are expected to exhibit a stronger second-order response, and consequently are especially interesting in studies of even-order nonlinear optical effects~\cite{Klein2006, Czaplicki2018, Huttunen2018b}.

\begin{figure*}[tbp]
\centering
{\includegraphics[width=0.80\linewidth]{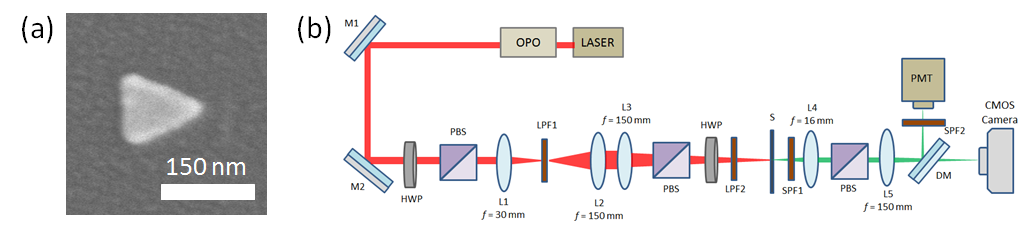}}
\caption{(a) Representative scanning electron micrograph of a fabricated gold nanoprism. (b) Schematic of the nonlinear SHG experimental setup (details in~\cite{Czaplicki2018}).
\label{fig:SEMandT}}
\end{figure*}

In this work, we study elongated triangular nanoprisms due to their simple geometrical shapes. Four metasurfaces containing gold nanoprisms with the widths $w=100$~nm, thicknesses $h=20$~nm and varying lengths of $l=145$, 156, 167 and 178~nm were fabricated on a fused silica substrate using electron beam lithography, thermal evaporation, and a standard metal lift-off procedure~\cite{awan2018fabrication}. 
In order to minimize inter-particle coupling effects occurring in periodic arrays~\cite{Auguie2008, Meinzer2014, kataja2015surface, Huttunen2016a, Kravets2018}, each metasurface consisted of 10,000 identical, randomly positioned nanoprisms (oriented in the same direction) that were deposited into an area of $200\times200$~$\mu$m$^2$. 
This allowed us to investigate ensemble responses that have spectral features identical to the responses of individual meta-atoms~\cite{Shi2014}. The arrangement is schematically represented in Fig.~\ref{fig:RLCschematic}(b), while a representative scanning electron micrograph of an individual nanoprism ($l=145$~nm) is shown in Fig.~\ref{fig:SEMandT}(a). Alternatively, individual meta-atoms could be investigated by using nonlinear microscopy~\cite{Bautista2012}.

\begin{figure*}[tbp]
\centering
{\includegraphics[width=0.40\linewidth]{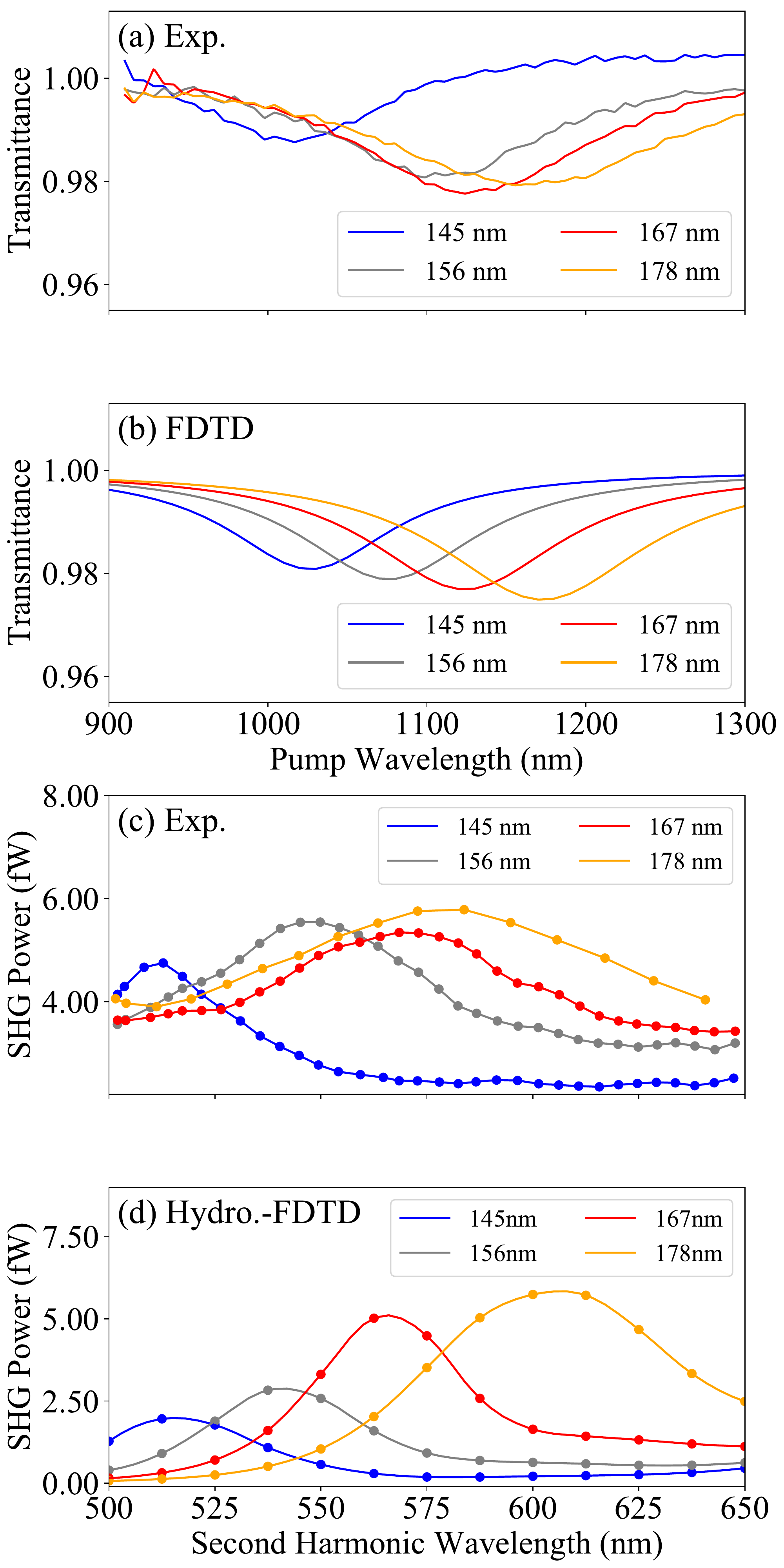}}
\caption{(a) Measured transmission spectra of fabricated random arrays of elongated gold nanoprisms of different lengths $l$ along the direction of the light polarization towards $x$-axis ($l=145$, 156, 167 and 178~nm). (b) FDTD calculations of the transmission spectra of random arrays of gold nanoprisms of different lengths. (c) Detected and (d) calculated SHG emission intensity as a function of incident wavelength for the four metasurface samples consisting of nanoprisms of varying lengths ($l=145$, 156, 167 and 178~nm).
\label{fig:Plots}}
\end{figure*}

After the samples were fabricated, we measured the transmission spectra of the four metasurfaces [see Fig.~\ref{fig:Plots}(a)]. To verify successful fabrication of the metasurfaces, we compared the measured spectra with the ones obtained from FDTD simulations (discussed in the Methods section) [see Fig.~\ref{fig:Plots}(b)]. In order for the simulated nanoprism to closely resemble the actual fabricated meta-atoms, we rounded its corners with a circle of 15-nm radius. 
The LSP resonances of simulated nanoprisms for $l=145$, 156, 167 and 178~nm peaked at 1030~nm, 1080~nm, 1120~nm and at 1170~nm, respectively, which are in good agreement with the measured resonance peak positions.

We performed SHG experiments using the setup shown in Fig.~\ref{fig:SEMandT}(b) (described in detail elsewhere~\cite{Czaplicki2018}). A laser beam originating from an optical parametric oscillator (Chameleon Compact) was used to illuminate the sample metasurfaces. The optical parametric oscillator was pumped with a Ti:Sapphire laser (Chameleon Vision II), generating 200-fs-long pulses with a repetition rate of 82~MHz. The average power of the signal beam was kept at 8~mW to avoid potential sample damage via accumulative heating. The SHG emission from the metasurfaces was detected as a function of fundamental wavelength ranging from 1000--1300~nm using a power-calibrated photomultiplier tube~[see Fig.~\ref{fig:Plots}(c)]. The input-beam polarization was set to be linear and aligned with the long axes of the nanoprisms.

After completing the SHG measurements, we verified the calibration of our setup to provide order-of-magnitude estimates for the first-order hyperpolarizabilities~$\beta$. This was achieved by measuring SHG emission from a 0.5-mm-thick Y-cut quartz crystal and using the model described in Ref.~[\citenum{Hayden1995}] to estimate the second-order susceptibility $\chi^{(2)}$ value for the quartz crystal. Our estimate ($\chi^{(2)}_{xxx}=0.53$~pm/V) was found to be in excellent agreement with the literature values, thereby verifying the calibration~\cite{Boyd2020}.

\begin{figure}[h]
\centering
{\includegraphics[width=\linewidth]{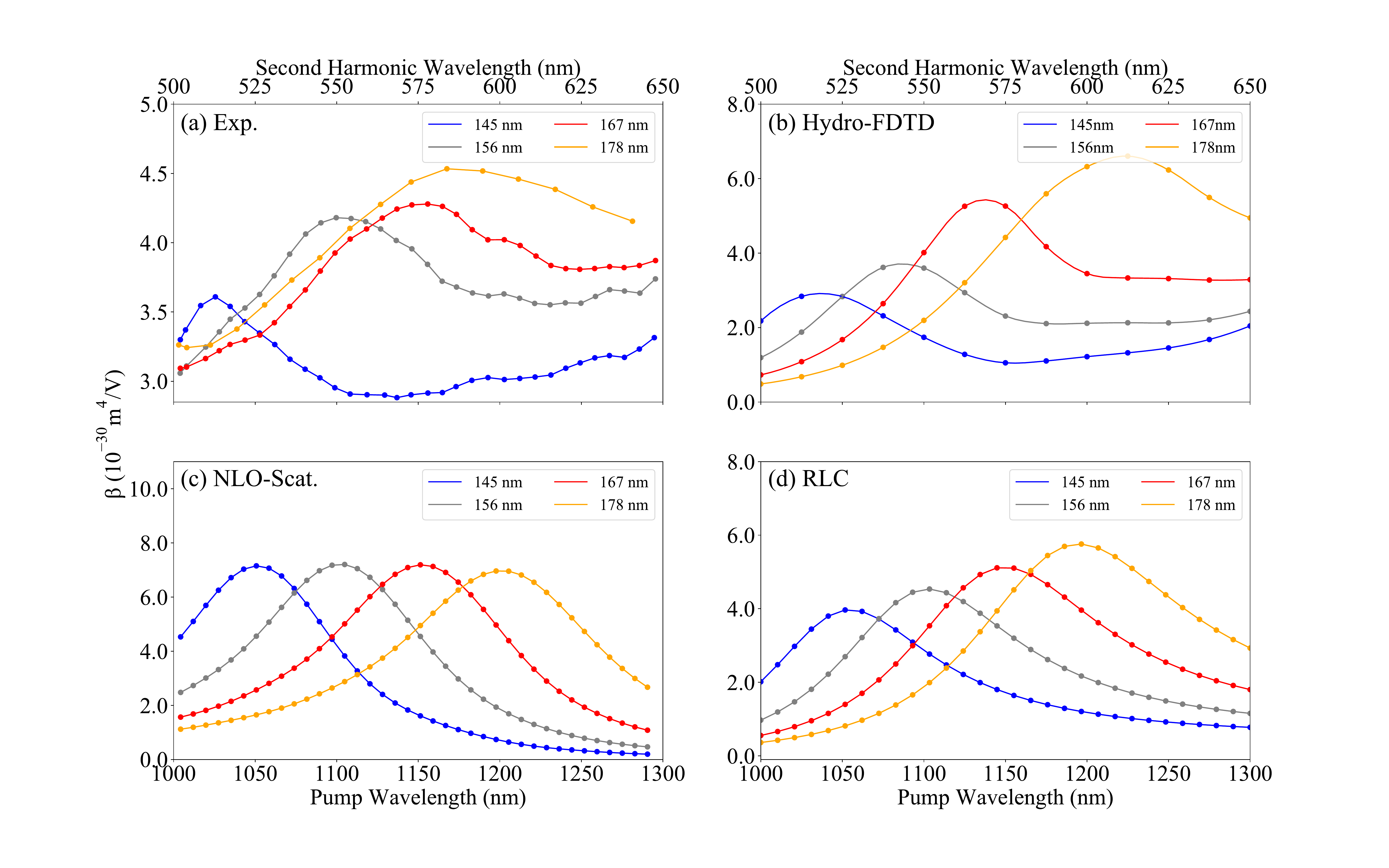}} 
\caption{(a) Experimentally extracted values of $\beta$ for the investigated nanoprism samples. The values of $\beta$, predicted by (b) Hydrodynamic-FDTD calculations, (c) the nonlinear scattering theory  and (d) the introduced nonlinear $RLC$ model.}
\label{fig:SHG_beta_data}
\end{figure}

To estimate the values of $\beta$ of the nanoprisms from the experimental data, we calculated the macroscopic $\chi^{(2)}$ values for the four metasurfaces using the same approach as we used when performing the reference measurements in quartz~\cite{Hayden1995}. We estimated the particle number density of the metasurfaces to be $n=10~000/(200~\mu$m$~\times~200~\mu$m$~\times~20~$nm$)= 1.25\times10^{19}~$m$^{-3}$,  then linked the detected SHG intensities to the hyperpolarizabilities using the relation $\beta = \chi^{(2)} / n$. The extracted values of $\beta$ as a function of the incident fundamental wavelength are plotted for the four investigated nanoprisms in Fig.~\ref{fig:SHG_beta_data}(a).

To further validate the experimental results, we used nonlinear hydrodynamic-FDTD simulations to calculate the SHG power and hyperpolarizability~$\beta$. The hydrodynamic plasma model~\cite{Sipe1980,Scalora2010,Ciraci2012,Ginzburg2015} was used in conjunction with the two-critical-points model~\cite{prokopidis_unified_2013} for gold in an in-house 3D-FDTD code~\cite{lesina_convergence_2015}. As the metasurface contains randomly positioned nanostructures, periodic boundary conditions could not be used to simplify the calculation. Further, the metasurfaces were too large for a single FDTD simulation. We therefore developed a new method for quantitative approximation of the power. The nonlinear scattered power from the metasurface was determined by first calculating the nonlinear scattered power in the forward direction of a single nanoprism from a Gaussian femtosecond pulse whose duration peak intensity matched the laser specification from the experiments. For each nanoprism in the random metasurface, the forward-scattered power was scaled to the incident field amplitude ``felt'' by that particle. The individual powers were then incoherently summed to estimate the total forward-emitted nonlinear power of the metasurface. These calculations were repeated for a number of pulses with central wavelengths ranging between 1000 and 1300~nm to generate the second-harmonic power spectrum, shown in Fig.~\ref{fig:Plots}(d). We see excellent quantitative agreement between the FDTD simulations and the experimental measurements.

We now estimate the values of $\beta$ of the nanoprisms using the hydrodynamic plasma model. We first calculate the nonlinear scattered power from a single nanoprism. This is done by integrating the nonlinear scattering spectrum from a single pulse and multiplying it by the repetition rate of the laser. If we assume that the nonlinear scattered power $P_\text{NL}$ is purely from a dipole, we can approximate the nonlinear dipole moment by $|p|^2=12\pi P_\text{NL}/(n c_0^2Z_0k_0^4)$~\cite{Jackson1999}, where $|p|$ is the magnitude of the nonlinear dipole moment, $c_0$ is the speed of light in vacuum, $k_0$ is the wave number in vacuum, $Z_0$ is the vacuum impedance, and $n$ is the refractive index of the surrounding medium. We calculate $\beta$ using Eq.~(\ref{Eq:polarizability}) and plot it in Fig.~\ref{fig:SHG_beta_data}(b). One can see excellent quantitative and qualitative agreement between the simulation and experimental results.

As the next step, we use another numerical approach based on the nonlinear scattering theory~\cite{Beer2011, Maekitalo2013} to calculate $\beta$. Though not as rigorous as the hydrodynamic plasma model, nonlinear scattering theory is a simpler and more computationally efficient technique for calculating nonlinear emission from nanostructures. Because the SHG emission was detected only in one direction, and the detector was essentially in the far-field zone of the source of radiation, we were able to simplify the simulations by making use of the Lorentz reciprocity theorem~\cite{Roke2004, Obrien2015}. First, we obtained the local field distributions for both the fundamental and SHG wavelengths of interest via linear simulations. The excitation field was assumed to be a normally incident plane wave polarized along the long axis of the nanoprism. The dimensions of the simulated nanoprism were matched with the experimental values, and the sharp corners of the meta-atoms were rounded to mimic the actual shapes of the fabricated meta-atoms. 

Then we calculated the generated nonlinear source polarization present on the surface of the meta-atom.  In the calculation, we used the experimentally extracted local nonlinear susceptibility values for gold~\cite{Wang2009}. 
The local field was transformed into the surface coordinate system by reconstructing the surface of the meta-atom by using a Delaunay triangulation mesh and performing subsequent field interpolation onto that surface.  Once the nonlinear surface polarization was calculated, we used the Lorentz reciprocity theorem to predict the emitted SHG field from the meta-atom (in the forward direction) by calculating the mode overlap integral between the SHG source polarization and the local field distribution at the SHG wavelength~\cite{Roke2009, Obrien2015}. The last step was to use Eq.~(\ref{Eq:polarizability}) to calculate $\beta$. The values of $\beta$ are plotted in Fig.~\ref{fig:SHG_beta_data}(c), and agree in resonance position and in the order-of-magnitude values with the experimental and hydrodynamic plasma model results.

Now that we have obtained values of $\beta$ from the experimental data and the two numerical approaches, we can estimate the value of the nonlinear coefficient $a$ from Eq.~(\ref{Eq:beta}). We find that $a\approx 10^{14}~\mathrm{C}^{-1}$, and plot Eq.~(\ref{Eq:beta}) in Fig.~\ref{fig:SHG_beta_data}(d). 

\begin{figure}[h]
\centering
{\includegraphics[width=0.40\linewidth]{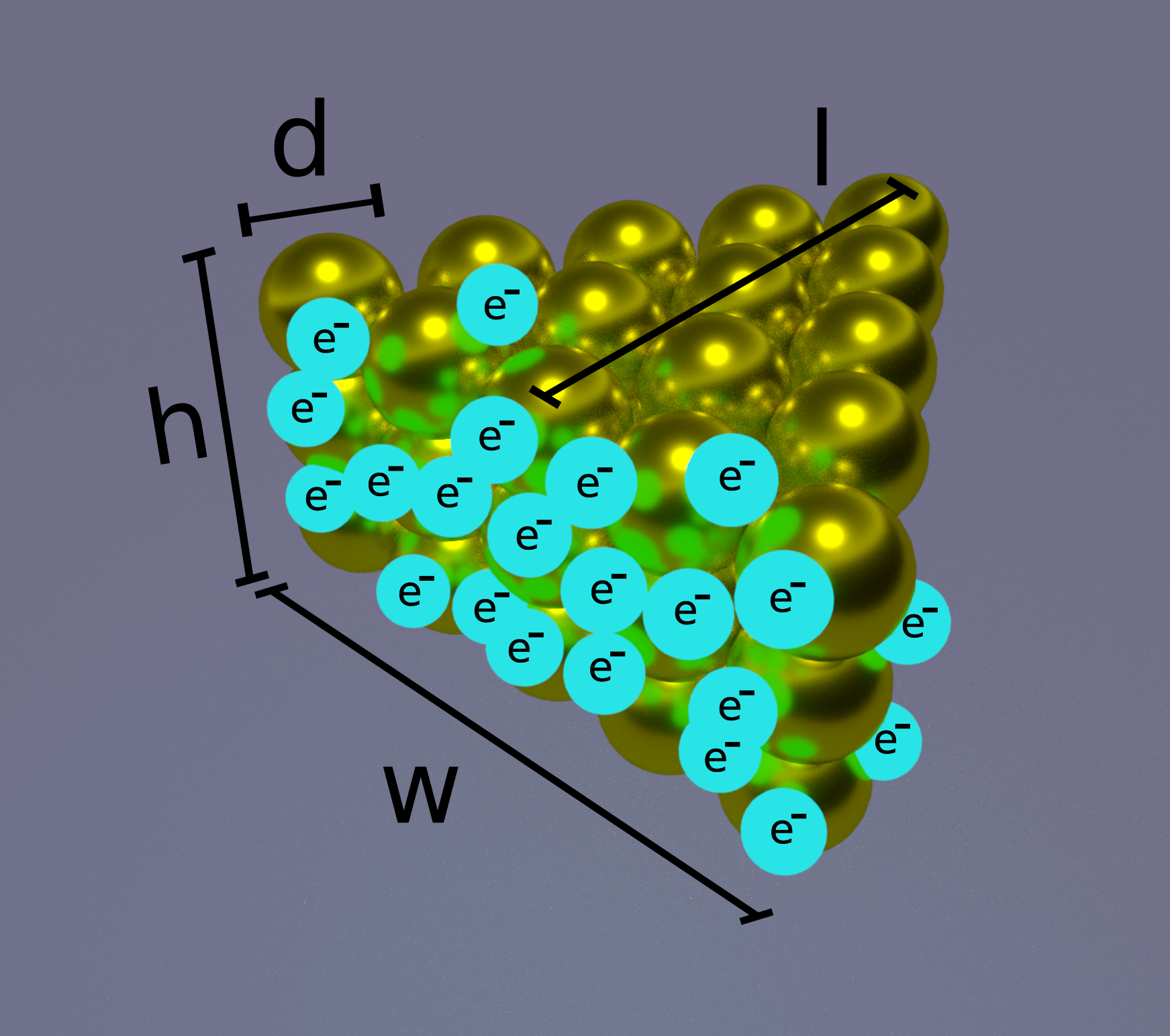}} 
\caption{Free-electrons are accumulated at the gold nanoprism's side surface (one atomic layer of thickness $d$) whose area is $A_{\mathrm{surf}} = w \times h$.}
\label{fig:Electron}
\end{figure}

While we are confident in the order of magnitude of this value of the nonlinear coefficient $a$, it was quite labour intensive to obtain. We now present a physically intuitive derivation for $a$ by applying a modified version of Miller's rule.~\cite{Boyd2020}. According to Miller's rule, the linear and nonlinear restoring forces, felt by an electron in a bulk material, will be comparable when the charge displacement is approximately equal to the size of an atom $d$ from which one can obtain a rough approximation for the nonlinear coefficient~\cite{Boyd2020}.
Here we are not dealing with a bulk material, but with a single meta-atom, where the perturbation that we are tracking within the $RLC$ model is the free-charge perturbation at the surface layer of the meta-atom, $\tilde{q}(t)$. In this case, a crude estimate of the nonlinear coefficient may be found by assuming that the linear and nonlinear components of the induced voltage, given by Eq.~(\ref{Eq:V_C}), are approximately equal to each other when all the free charge contained within one atomic layer accumulates at the surface, as illustrated in Fig.~\ref{fig:SHG_beta_data}. We call this the equilibrium free charge, and it is given by
 \begin{equation}  
 \label{Eq:Miller3}
  q_\mathrm{eq} = A_\mathrm{surf}d\rho_\mathrm{free}  \, ,
 \end{equation}
where $A_\mathrm{surf}$ = 100~nm$\times$20~nm is the maximum surface area of the side wall of the nanoprism opposite to the acute angle in Fig.~\ref{fig:Electron} (where the charge builds up), and $d = 0.41$~nm is the lattice constant of gold. The free charge density  of gold, $\rho_\mathrm{free}$, can be expressed as
 \begin{equation}  
 \label{Eq:Miller4}
  \rho_\mathrm{free} = -e n_\mathrm{free}  \, ,
 \end{equation}
where $n_\mathrm{free} = 5.9 \times {10}^{28}$~m$^{-3}$ is the free electron number density of gold, and $e$ is the charge of an electron.

Setting the linear and nonlinear components of $V_\mathrm{C}$ in Eq.~(\ref{Eq:V_C}) to be approximately equal, we have
 \begin{equation}  
 \label{Eq:Miller5}
  q_\mathrm{eq} = aq_\mathrm{eq}^2  \, ,
 \end{equation}
which yields
 \begin{equation}  
 \label{Eq:Miller5}
  a = \frac{1}{q_\mathrm{eq}} = \frac{-1}{d A_\mathrm{surf} e n_\mathrm{free}} = -1.3 \times {10}^{14}~\mathrm{C}^{-1} \, .
 \end{equation}
We thus have obtained a physically intuitive formula for $a$ that gives the correct order of magnitude estimate, and is in excellent agreement with experimental and both sets of simulation results. This procedure to estimate $a$ is valid for any bar-like nanostructure.

From Fig.~\ref{fig:SHG_beta_data}, we see that all four approaches yield values of $\beta$ 
within the same order of magnitude ${\sim}10^{-30}~$m$^4$/V. This agreement is very encouraging because it has been notoriously difficult to make quantitative predictions of nonlinear optical processes occurring in plasmonic materials~\cite{ButetReview2015, Czaplicki2018}. It is worth pointing out that the simpler methods ($RLC$ and nonlinear scattering theory), though predicting the correct spectral peak positions, are missing some features that are visible in the experiment and hydrodynamic calculation, such as the oscillations at longer wavelengths. These features are believed to be caused by a secondary resonance of the nanoprisms and interband transitions in gold.  

Despite the great amount of previous work on nonlinear metasurfaces, only a handful of investigations have provided order-of-magnitude estimates of meta-atom's hyperpolarizabilities~\cite{Vance1998,RussierAntoine2004, Nappa2005,Duboisset2009, Butet2010}. We note here that our calculated values of hyperpolarizability are 2--3 orders of magnitude smaller compared to the previous estimates of somewhat similar meta-atoms. However, this discrepancy is not unexpected, because the experimental setups, the wavelength ranges considered, and the investigated meta-atoms have all been different in these studies. 
Here, we have investigated the coherent SHG emission from meta-atoms, whereas earlier investigations have measured incoherent hyper-Rayleigh scattering (HRS) signals, and have estimated the values of $\beta$ indirectly by comparing the HRS signals from meta-atoms with HRS signals measured from known solvents. Furthermore, the earlier HRS experiments have been performed at shorter excitation wavelengths than what we have used in our study. Because the interband transitions of gold start playing a role at wavelengths shorter than 550~nm~\cite{novotny2006book}, these earlier-extracted hyperpolarizability values may have contained an additional contribution arising from the inter-band transitions~\cite{Boyd2020}.

In addition to the consistent calculations for the hyperpolarizabilities, our results also demonstrate the usefulness of the nonlinear $RLC$ approach. Although the approach is simple and intuitive, it appears to describe the dynamics of the conduction electrons adequately enough to allow predicting the strength of nonlinear optical responses of plasmonic meta-atoms. The nonlinear $RLC$ model has earlier been found to accurately describe nonlinear responses of metamaterials at the microwave wavelengths~\cite{Poutrina2010}. Here, we show that the $RLC$ approach can be adapted for optical wavelengths 
by properly taking into account the plasmonic behavior of metals.

Our simple $RLC$ framework also makes it clear that the optical nonlinearity is directly linked to the coherent dynamics of the conduction electrons, prompting more detailed investigations on that matter. This is further justified by the excellent agreement between the free-electron hydrodynamic calculations and the experimental measurements. One can thus conclude that the proposed $RLC$ model can be used to predict both linear and nonlinear optical responses of meta-atoms.

To conclude, we have demonstrated that the nonlinear equivalent $RLC$ circuit model can be used to quickly and accurately predict the nonlinear optical responses of meta-atoms in the visible and near-IR spectral ranges. We fabricated four metasurfaces consisting of randomly positioned gold nanoprisms and characterized their second-harmonic generation emissions. We compared the experimental results with the predictions based on the hydrodynamic plasma model, nonlinear scattering theory, and the introduced nonlinear $RLC$ model. All the results were found to be in good agreement. Our $RLC$ approach provides new insights into understanding the nonlinear responses of meta-atoms and opens new possibilities for quickly designing nonlinear metasurfaces.



\section*{Methods}

\subsection*{Parameters for the Nonlinear $RLC$ Model}
\textbf{Formula describing the capacitance $C$ and the inductance $L$ of an elongated bar-like plasmonic nanostructure}: The capacitance $C$ and inductance $L$ of an elongated nanoprism can be calculated from its geometrical dimensions using equations
\begin{subequations} 
\begin{eqnarray}  
C &=& 2\pi \epsilon_{0}\epsilon_{r} \frac{h}{2} \, , \qquad  \\ 
\label{Eq:Capacitance_rect}
L &=& \frac{\mu_{0} l} {2 \pi} \log \frac{4 l}{h} + \frac{\mu_{0} l}{\frac{{\omega_\mathrm{p}}^2}{{c_0}^2} w h}\, ,  
\label{Eq:Inductance_rect}
\end{eqnarray}
\end{subequations}

where the self-inductance is $L_\mathrm{self} = \frac{\mu_{0} l} {2 \pi} \log \frac{4 l}{h}$, the kinetic-inductance is $L_\mathrm{kinetic} = \frac{\mu_{0} l}{\frac{{\omega_\mathrm{p}}^2}{{c_0}^2} w h}$, $\epsilon_{0}$ is the vacuum permittivity ($8.85 \times 10^{-12} $~F/m), the relative permittivity of glass is $\epsilon_{r} \approx 3.9$, $\mu_{0}$ is the vacuum permeability ($1.257 \times 10^{-6} $~H/m), $\omega_p = 13.8 \times 10^{15}$~rad/s is the plasma frequency for gold in the optical regime~\cite{Johnson1972} 
and $c_0 = 3 \times 10^{8}$~m/s is the speed of light in vacuum. A similar model was proposed for a plasmonic cylindrical nanorod in Ref.~[\citenum{Huang2009}]. Due to the geometrical difference of our nanostructures, our model is based on modified equations for the capacitance and inductance. 

\subsection*{Simulations}


\textbf{Linear FDTD Simulations:} Linear FDTD calculations were performed to simulate the linear transmission spectra of the metasurface using an in-house FDTD solver~\cite{lesina_convergence_2015}. The spectra was calculated by subtracting the absorption and back-scattering cross-sections of all the meta-atoms on the metasurface from the total area of the metasurface. The transmission spectrum is then: $T=(A_\mathrm{metasurface}-(A_\mathrm{back-scatt}+A_\mathrm{abs}))/A_\mathrm{metasurface}$. The cross-sectional data is calculated from a single meta-atom and scaled by the number of meta-atoms on the metasurface. A standard total-field/scattered-field layout is used to calculate the cross-sections and the simulation domain is truncated by convolutional perfectly matched layers. The linear Drude + 2 critical points model~\cite{prokopidis_unified_2013} is used for the optical properties of gold and accounts for contributions from the conduction electrons and interband transitions. A broadband raised cosine pulse \cite{lesina_convergence_2015} is used as a source excitation.

\noindent \textbf{Hydrodynamic FDTD Simulations:} Hydrodynamic FDTD calculations were conducted using the same in-house FDTD solver. The simulation setup is identical to that used in the linear transmission spectra, except that the hydrodynamic model (solved via centered finite differences) replaces the Drude model (in the Drude + 2 critical points model) and the source excitation is replaced with a 200 fs Gaussian pulse centered at wavelengths ranging from 1000 nm - 1300 nm. All simulations using the in-house FDTD solver were run on the Graham cluster operated by Compute Canada.

\noindent \textbf{Nonlinear Scattering Theory:}
The nonlinear response of a meta-atom was estimated also by using calculations based on the nonlinear scattering theory and the Lorentz reciprocity theorem~\cite{Roke2009, Obrien2015}. The strength of the SHG emission in the direction of interest was evaluated by calculating a mode overlap integral over the fundamental excitation and SHG emission modes. The relevant field profiles for the fundamental and SHG fields were calculated using the FDTD method, and the mode overlap integrals were calculated numerically using Matlab. In the FDTD simulations the optical constant of gold was taken from~[\citenum{Johnson1972}]. The fields on the surface of the meta-atom were estimated by using Delaunay triangulation, and only the surface contributions were considered when calculating the nonlinear response of the gold~\cite{Wang2009}. 

\subsection*{Fabrication}
We used 2~cm~$\times$~2~cm fused silica chips as substrates. The chips were coated with bi-layer electron-beam resist, consisting of 50-nm-thick PMMA with a molecular weight of 495~k as the bottom layer and 25-nm-thick PMMA with a molecular weight of 950~k as the top resist layer. The plasmonic nanostructures were then patterned using 30-kV Raith electron-beam  lithography system (CRPuO, uOttawa) with a dose of 550 $\mu$C/cm$^2$. The patterned resist was then developed for 2 minutes in 3:1 MIBK:IPA (Methyl isobutyl ketone-Isopropyl Alcohol), followed by depositing a 20-nm layer of gold by electron-beam evaporation and, finally, a lift-off by immersion in acetone. A computer-aided layout of a randomly arranged elongated nanoprism array that was used to create the resist mask and a schematic of the fabrication process flow are shown in Fig.~ \ref{FIG:S2} and Fig.~ \ref{FIG:S3}, respectively, in the supplementary information S2 and S3. More details are described in Ref.~[\citenum{awan2018fabrication}].

\subsection*{Characterization}
\textbf{Linear Characterization:} Linear transmission spectra of the samples were measured using  a collimated tungsten-halogen light source (experimental setup is shown in Fig.~\ref{FIG:S4} in the supplementary information S4). The incident polarization was controlled using a broadband linear polarizing filter. The entire sample was illuminated, and the transmission from a single device was measured by first using a lens to image the sample plane into an intermediate image plane. The transmission from the correct device was then selected by a translating a variable aperture in this image plane, and by using a second lens to guide the transmitted light into the spectrometer.  

\textbf{Nonlinear Characterization:} Signal beam from an optical parametric oscillator (Chameleon
Compact) was used to illuminate the sample metasurfaces using a spectral SHG setup described in the supplementary information S5 (shown in Fig.~\ref{FIG:S5}). The optical parametric oscillator was pumped using a Ti:Sapphire laser (Chameleon Vision II) generating 200-fs-long pulses with a repetition rate of 82~MHz. The average power of the signal beam was kept at 8~mW to avoid potential sample damage via accumulative heating. The SHG emission from the metasurfaces was detected as a function of the fundamental wavelength ranging between 1000 and 1300~nm using a power-calibrated photomultiplier tube~(see Fig.~ \ref{FIG:S5} in the supplementary information S5). The input beam polarization was set to be linear and aligned with the long axes of the nanoprisms. 

\section*{Author contributions}
KD conceived the basic idea for the whole work. KLT conceived the idea of implementing the $RLC$ model. MSBA derived the analytical nonlinear $RLC$ model. KMA fabricated the metasurface substrate. MJH, MSBA, and AK carried out the measurements. JB performed the linear and nonlinear Hydrodynamic-FDTD model simulations. JB and LR dervied the model for the nonlinear coefficient \textit{a} in the spirit of Miller's rule. YM made the illustration. KD, LR, MJH, KT, and ACL supervised the research and the development of the manuscript. MSBA, JB, and MJH wrote the first draft of the manuscript. All co-authors subsequently took part in the revision process and approved the final copy of the manuscript.

\begin{acknowledgement}

The authors thank Robert W. Boyd, Ekaterina Poutrina, John E. Sipe, and Gerd Leuchs for their valuable feedback and suggestions. The authors also thank Martti Kauranen for providing his lab at Tampere University in Finland to perform the experiment. 

KD and LR acknowledge support from the Canada Research Chairs (CRC) Program. MSBA acknowledges the support of the Ontario Graduate Scholarship (OGS), the University of Ottawa Excellence Scholarship, and the University of Ottawa International Experience Scholarship. JB acknowledges the financial support of the Natural Sciences and Engineering Research Council of Canada (NSERC) Canada Graduate Scholarship-Master’s (CGSM) Program and the University of Ottawa Excellence Scholarship, and the computational resources of Compute Canada. MJH acknowledges the support of the Academy of Finland (Grant No. 308596) and the Flagship of Photonics Research and Innovation (PREIN) funded by the Academy of Finland (Grant No. 320165). KLT acknowledges support from the General Secretariat for Research and Technology(GSRT) and the Hellenic Foundation for Research and Innovation (HFRI) under Grant 1819. 

\end{acknowledgement}





\providecommand{\latin}[1]{#1}
\providecommand*\mcitethebibliography{\thebibliography}
\csname @ifundefined\endcsname{endmcitethebibliography}
  {\let\endmcitethebibliography\endthebibliography}{}



\clearpage

\onecolumn
\renewcommand\thepage{S\arabic{page}} 
\setcounter{page}{1}
\renewcommand\thesection{S\arabic{section}} 
\setcounter{section}{0}
\renewcommand\thefigure{S\arabic{figure}}   
\setcounter{figure}{0}  
\renewcommand\theequation{S\arabic{equation}} 
\setcounter{equation}{0}



{\Huge Supplementary Information}
\vspace{1em}

\begin{figure}[H]
    \centering
   \includegraphics[width=1\linewidth]{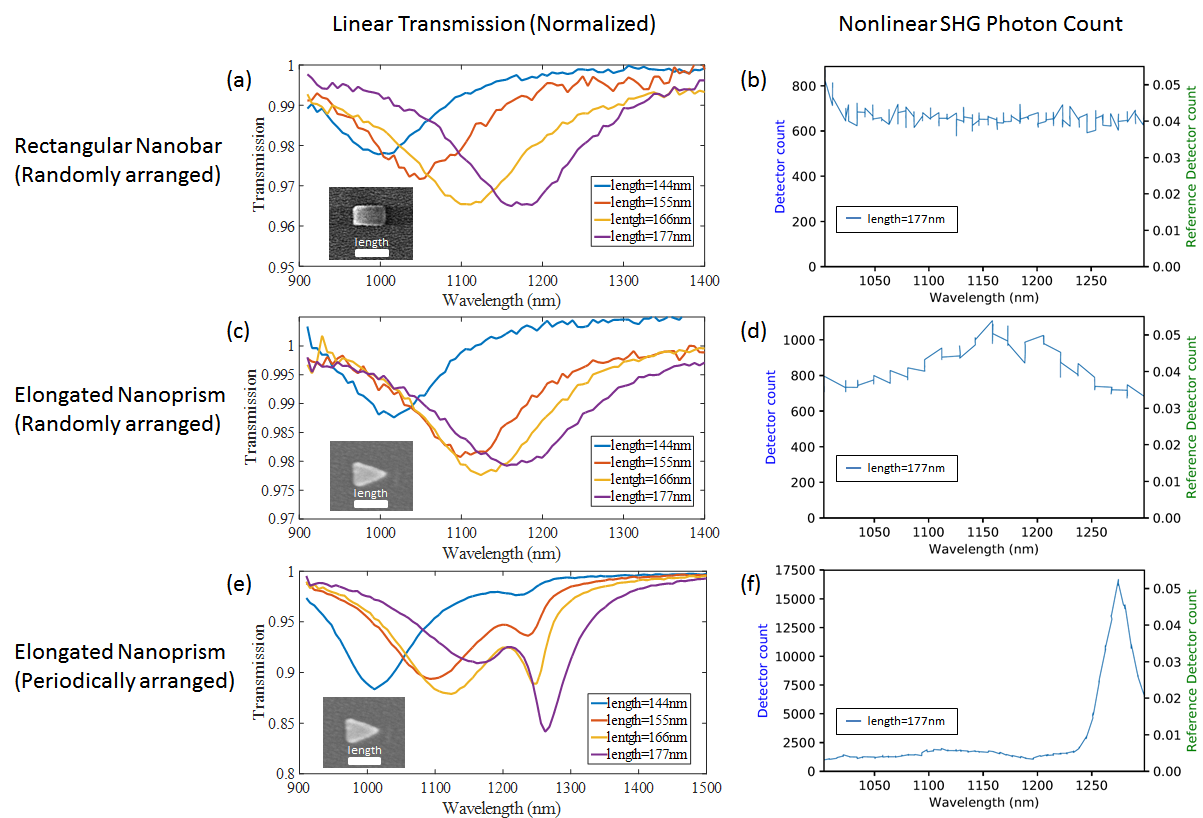} 
    \caption{{\bf~|~Dependence of SHG on Nanostructures' Symmetric and Non-Symmetric shapes, and the random and periodic lattice arrangement.} Linear transmission and SHG photoncount as a function of incident wavelength. (a-b) Rectangular nanobars arranged randomly, (c-d) elongated nanoprisms arranged randomly, and (e-f) elongated nanoprisms arranged periodically in the metasurface arrays.}
    \label{FIG:S1}
\end{figure}

\begin{figure}[H]
\centering
\includegraphics[width=1\linewidth]{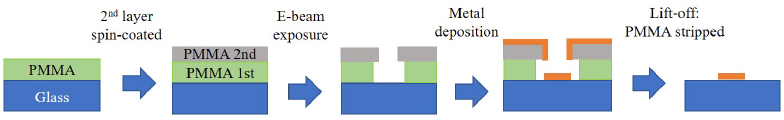}
\caption{{\bf~|~Fabrication Process.} Schematic of the fabrication process flow.}
\label{FIG:S3}
\end{figure}

\noindent Below is the supplementary information for \emph{Hyperpolarizability of plasmonic meta-atoms in metasurfaces} by M. Saad Bin-Alam, Joshua~Baxter, Kashif M.~Awan, Antti~Kiviniemi, Yaryna~Mamchur,  Antonio Cal\`a Lesina, Kosmas L.~Tsakmakidis, Mikko J. Huttunen, Lora Ramunno, and Ksenia Dolgaleva. In Sec.~\ref{SEC:NonSymm}S1, Fig.~\ref{FIG:S1} shows a comparison between the symmetric and non-symmetric nanostructures' SHG response. In Sec.~\ref{SEC:CAD}S2, Fig.~\ref{FIG:S2} shows a computer-aided design of the layout of the randomly oriented elongated nanoprisms prior to the Electron-beam lithography process of the plasmonic metasurfaces. In Sec.~\ref{SEC:Fab}S3, Fig.~\ref{FIG:S3} shows a schematic of the fabrication process of the plasmonic metasurfaces.  In Sec.~\ref{SEC:Linear_Setup}S4, Fig.~\ref{FIG:S4} shows the experimental setup (a short description is also provided) we used to measure the linear transmittance of the fabricated metasurfaces. In Sec.~\ref{SEC:SHG_Setup}S5, we elaborately describe the details of the experimental setup (shown in Fig.~\ref{FIG:S5}) used in SHG power measurement.

\section{\textbf{S1}: Effect of the Symmetric and Non-Symmetric Shapes, and the lattice arrangement of the Nanostructures on Second-Harmonic Generation (SHG)}\label{SEC:NonSymm}

\section{\textbf{S2}: Computer-Aided Design (CAD) Layout of the Array of the Randomly Arranged Elongated Nanoprism}\label{SEC:CAD}

\begin{figure}[H]
\centering
\includegraphics[width=1\linewidth]{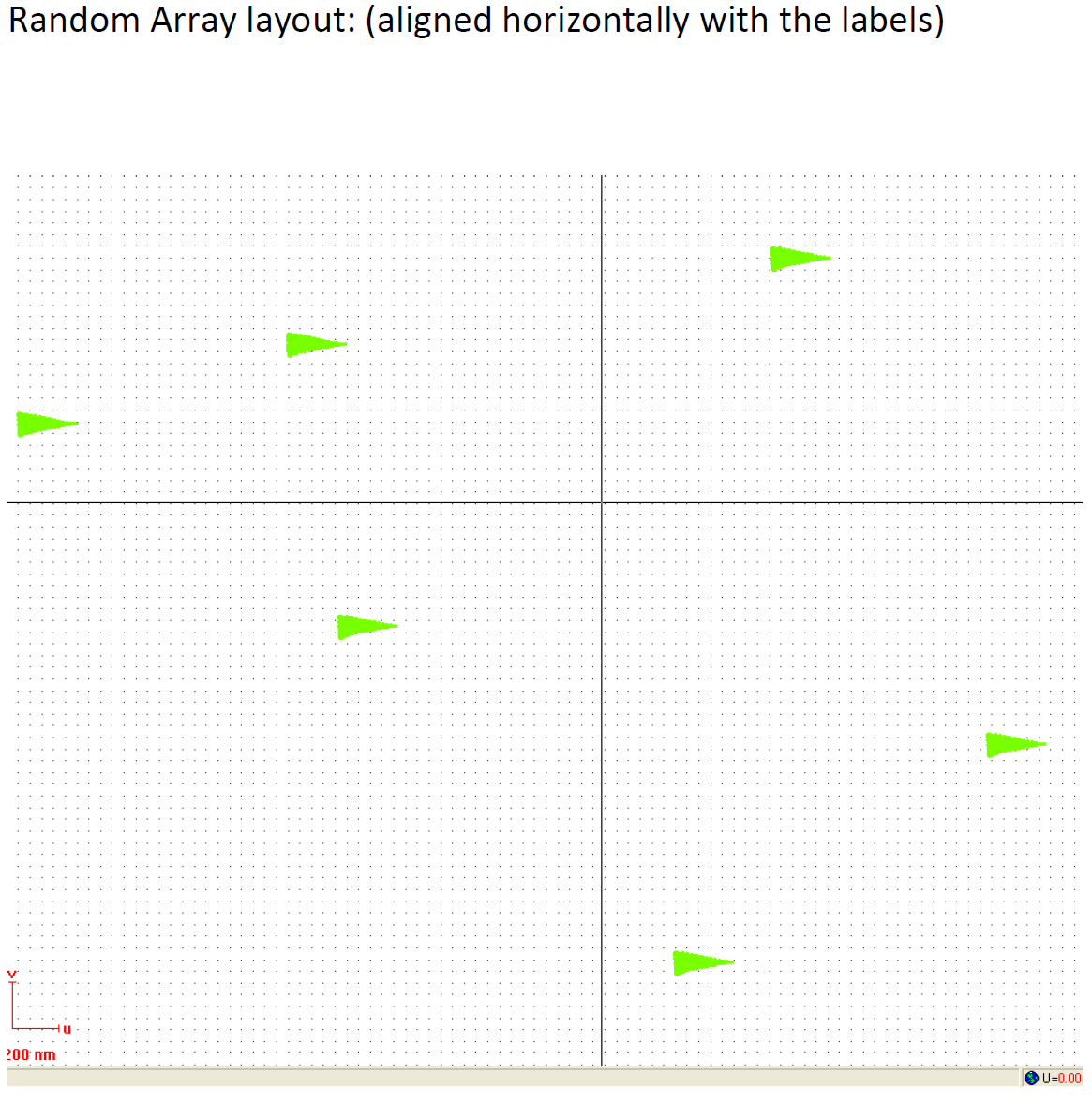}
\caption{{\bf~|~Computer-Aided Design (CAD) Image.} A computer-aided layout of a randomly arranged elongated nanoprism array that is used to create the resist mask in the Electron-beam lithography.}
\label{FIG:S2}
\end{figure}




\section{\textbf{S3}: Plasmonic Metasurface Fabrication Process}\label{SEC:Fab}

\section{\textbf{S4}: Experimental Setup for Linear Transmittance Measurement} \label{SEC:Linear_Setup}
\begin{figure}[H]
\centering
\includegraphics[width=1\linewidth]{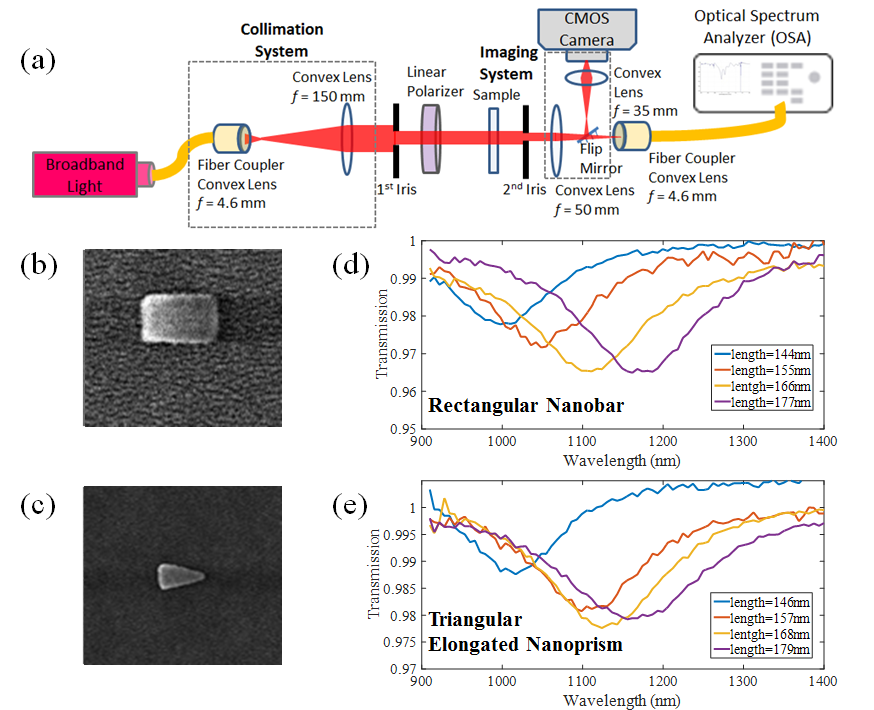}
\caption{{\bf~|~Linear Setup and Measurement.} (a) Experimental setup for the linear transmission measurement. (b-c) SEM images of a Rectangular nanobar and an elongated nanoprism. (d-e) corresponding transmission spectra of varying lengths (average lengths for both shapes: $l=145$, 156, 167 and 178~nm).}
\label{FIG:S4}
\end{figure}

Fig.~\ref{FIG:S4}(a) shows the schematic of the linear transmittance measurement setup. A broadband source is collimated and is polarized using a broadband linear polarizing filter. A first iris is optionally placed to help align the sample in the center of the beam. The beam is then passed through the sample. The surface of the device is imaged using a lens, and a pinhole is placed in the image plane to select the desired array. The transmitted light is collected in a large core multimode fiber and is analyzed using an optical spectrum analyzer. Fig.~\ref{FIG:S4}(b-c) shows the SEM images of a Rectangular nanobar and an elongated nanoprism. Fig.~\ref{FIG:S4}(d-e) shows the corresponding transmission spectra of varying lengths (average lengths for both shapes: $l=145$, 156, 167 and 178~nm)

\section{\textbf{S5}: Details of nonlinear experimental setup}\label{SEC:SHG_Setup}
A schematic of the SHG experimental setup is illustrated in Fig.~\ref{FIG:S2}. A motorized achromatic half-wave plate (HWP) and a polarizer were used to control the level of power $P_{\omega}$ from the OPO. Before entering the polarization-control part of the setup, the fundamental beam was cleaned and expanded with a set of lenses and an aperture (diameter $D = 25$~$\mu$m). To weakly focus the beam on the sample arrays, an achromatic lens with the focal length of 150~mm was used, ensuring a relatively small beam waist diameter of the excitation beam (around 100 $\mu$m) while the plane-wave approximation could still be used. To control the input polarization, we used a high-quality polarizer and an achromatic HWP, whereas to select the polarization of the emitted SHG light, we used a film polarizer after the sample. Then, to pass (block) the fundamental beam, we used a 900~nm long-pass (700~nm short-pass) filter. To efficiently collect the generated SH signal, a lens with the focal length of 16~mm was used after the sample. After being reflected by a dichroic mirror and passing through another short-pass filter (900~nm), the SHG signal was focused on the active area of a photomultiplier tube (PMT) module with another achromatic lens of 150~mm focal length. For sample alignment, the light transmitted through the dichroic mirror was used to image the sample plane with a CMOS camera and a camera lens (MVL50M23). The PMT has been calibrated using a sensitive power meter, and the result is that 1~count/s corresponds to $5.2$~aW.

\begin{figure}[H]
     \centering \includegraphics[width=1\linewidth]{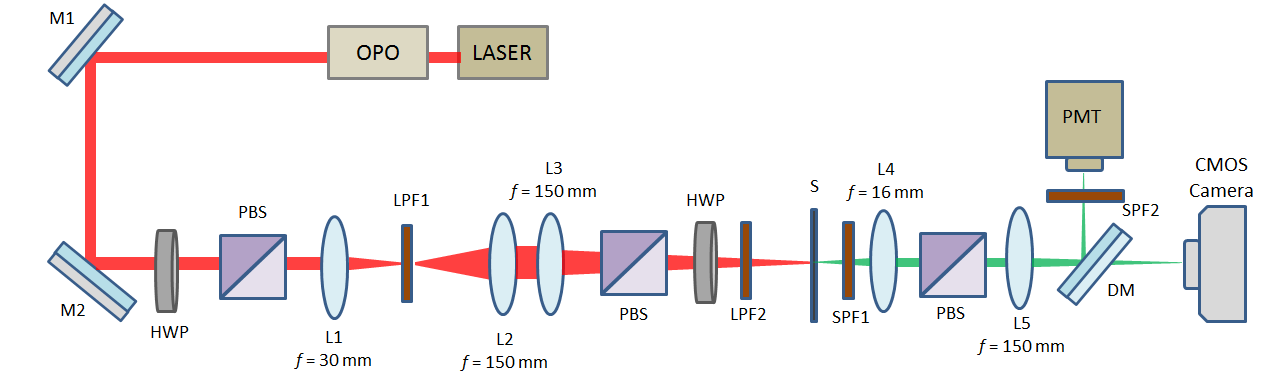}
     \caption{{\bf~|~Nonlinear SHG Experimental Setup.} Schematic representation of the experimental setup for measuring SHG. M – mirrors, HWP – motorized half-wave
plates, P – polarizers, S - sample, L – lenses (L1 has $f=30$~mm, L2 has $f=150$~mm, L3 has f=150 mm, L4 has f=16 mm, L5 has f=150 mm, L1, L2, L3, and L5 – achromats). LPF –- long-pass filter at 900~nm, SPF1
–- short-pass filter at 700~nm, A – film polarizer (analyzer), DM – dichroic mirror, SPF2 – short-pass filter at 900~nm, PMT – photomultiplier tube (PicoQuant PMA-C 192-M).}
     \label{FIG:S5}
\end{figure}





\end{document}